\documentclass[12pt,aps,prb,preprint]{revtex4}   % style for Physical Review B and AJP are similar

\usepackage{amsmath}    % need for subequations
\usepackage{graphicx}   % for figures

\begin{document}

\title{The Nature of the Vector and Scalar Potentials and Gauge Invariance in the Context of Gauge Theory}
\author{Tony Stein}
 \affiliation{Southwestern Oklahoma State University, Department of Chemistry and Physics, Weatherford, OK 73096}
 \email{tony.stein@swosu.edu}   %optional
\date{\today}

\begin{abstract}
Modern undergraduate textbooks in electricity and magnetism typically focus on a force representation of electrodynamics with an emphasis on Maxwell's Equations and the Lorentz Force Law.  The vector potential $\mathbf{A}$ and scalar potential $\Phi$ play a secondary role mainly as quantities used to calculate the electric and magnetic fields.  However, quantum mechanics including quantum electrodynamics (QED) and other gauge theories demands a potential ($\Phi$,$\mathbf{A}$) oriented representation where the potentials are the more fundamental quantities.  Here, we help bridge that gap by showing that the homogeneous Maxwell's equations together with the Lorentz Force Law can be derived from assuming that the potentials represent potential energy and momentum per unit charge.  Furthermore, we enumerate the additional assumptions that are needed to derive the inhomogeneous Maxwell's equations.  As part of this work we demonstrate the physical nature and importance of gauge invariance.    
\end{abstract}

\maketitle

\section{Introduction}

\noindent The vector and scalar potentials have an interesting history.\cite{History}  James Clerk Maxwell originally  formulated his equations using the vector potential $\mathbf{A}$ along with the electric $\mathbf{E}$ and magnetic $\mathbf{B}$ fields.  In his first great paper published in 1856, Maxwell showed that Michael Faraday's experimental work in electrodynamics could be expressed as $\mathbf{E}=-\frac{\partial}{\partial t}\mathbf{A}$ where $\mathbf{E}$ is the \emph{induced} electric field.  Since $\mathbf{E}$ is defined as the force per unit charge Maxwell deduced that $\mathbf{A}$ represents a \textit{potential momentum} per unit charge in the same way that the scalar potential $\Phi$ represents a \text{potential energy} per unit charge.

In 1885, Oliver Heaviside eliminated the vector potential from Maxwell's equations in favor of using $\mathbf{E}$ and $\mathbf{B}$ only.  Heaviside viewed the $\mathbf{E}$ and $\mathbf{B}$ fields as the true physical quantities with the potentials being merely useful functions.  This view was likely influenced by the gauge invariance of the potentials.  Adding the gradient of an arbitrary function to $\mathbf{A}$ left the fields the same provided you subtracted the time derivative of the same function from $\Phi$.

Forty years later the advent of quantum mechanics began to reassert $\mathbf{A}$ and $\Phi$ as important quantities in their own right.  In quantum mechanics, energy and canonical momentum are fundamental quantities. Force and consequently force fields have a much more limited role.  Further, gauge invariance was shown to be equivalent to the phase invariance inherent in quantum mechanics\cite{History,Weyl,Translation:Weyl}.  Later still, the Aharonov Bohm effect proved that the potentials have noticeable effects in the absence of electric and magnetic fields \cite{Ehrenberg&Siday,Aharonov&Bohm} and therefore established the potentials as physical quantities in their own right.  

The true importance of the vector and scalar potentials became most apparent with the development of quantum electrodynamics (QED) and subsequent gauge theories.  If we ask how the electromagnetic force gets from particle A to particle B then Maxwell had a simple answer.  Force is carried by electric and magnetic fields through an ether similar to how transverse sound waves move through a solid.  The  physical nature of $\mathbf{E}$ and $\mathbf{B}$ makes sense in this context as does the Maxwell stress tensor.  Unfortunately, ether does not exist.  In QED, the force is transfered not through an elastic medium but by particles, photons.  But, particles cannot carry force; particles carry (or have) \textit{energy and momentum}.  The $\mathbf{A}$ and the $\Phi$ fields represent how this momentum and energy is carried.  The force type fields $\mathbf{E}$ and $\mathbf{B}$ are the derived quantities.  (See E. J. Konopinski\cite{Konopinski} p. 502 or R. Feynman\cite{Feynman}.)  

Yet almost 60 years later, the nature of the potentials is often downplayed in favor of the electric and magnetic fields.  Recent work \cite{Konopinski,Semon&Taylor,Redzic,Adachi} has begun to revive the potentials as representing the potential energy and momentum per unit charge.  Some of this work has been incorporated into the undergraduate electrodynamics curriculum \cite{Griffiths}, but more needs to be done.\cite{Feynman}  Here we will extend that work by showing that the laws of electrodynamics result naturally from the physical meaning of the potentials in the context of the assumptions behind QED. 

\section{The Lorentz Force Law}

We start by deriving the Lorentz force on a particle with charge $q$ and moving with a speed $v$.  We assume that the net force on the particle due to all the other charges in the universe can be described entirely in terms of a potential energy per unit charge $\Phi$ and potential momentum per unit charge $\mathbf{A}$.  This assumption is known as \textit{minimal coupling}.  Minimal coupling is a reasonable assumption based on the fact that electromagnetic force is mediated by photons---which as particles carry energy and momentum. With minimal coupling, the total energy $H$ and momentum $\mathbf{P}$ of our charged particle moving at a speed $\mathbf{v}$ are:

\begin{eqnarray}
H = E + q\Phi, \nonumber \\ 
\mathbf{P} = \mathbf{p} + q\mathbf{A},
\label{eq:H,P}
\end{eqnarray}

\noindent where $E$ and $\mathbf{p}$ represent the relativistic energy and momentum of the charged particle.  

In special relativity the quantity 

\begin{equation}
  \left(\frac{E}{c}\right)^2 - \mathbf{p} \cdot \mathbf{p} = (mc)^2
\label{eq:p^2} 
\end{equation}

\noindent is invariant for all reference frames, where m is the rest mass of the particle and c is the speed of light in a vacuum.  Inserting Eq. (\ref{eq:H,P}) into Eq. (\ref{eq:p^2}) and solving for H in terms of P we obtain the Hamiltonian for the system:\cite{Jackson}
\begin{equation}
  H = c \sqrt{(\mathbf{P}-q\mathbf{A})^2+(mc)^2}+q\Phi.
  \label{eq:H(P,x)}
\end{equation}
\noindent The force $\mathbf{F}$ on the particle is equal to the time derivative of its momentum $\mathbf{F} = \dot{\mathbf{p}} = \dot{\mathbf{P}}-q\dot{\mathbf{A}}$.  We can therefore use Hamilton's equations of motion ($\dot{x} = \frac{\partial H}{\partial P_x}$ and $\dot{P}_x = -\frac{\partial H}{\partial x}$) to determine $\mathbf{F}$.  Assuming $\mathbf{A}$ and $\Phi$ have no explicit dependence on $\mathbf{P}$ and using $\dot{x} = v_x = \frac{\partial H}{\partial P_x} = -\frac{1}{q} \frac{\partial H}{\partial A_x}$ we obtain that
\begin{equation}
F_x = \left(\dot{\mathbf{p}}_q \right)_x = -q\dot{A}_x-\frac{\partial}{\partial x}\left(q\Phi - q \mathbf{v}\cdot\mathbf{A} \right).
\label{eq:F_x}
\end{equation}
\noindent Here we have also used $\mathbf{v}\cdot\frac{\partial\mathbf{A}}{\partial x} = \frac{\partial}{\partial x}(\mathbf{v}\cdot\mathbf{A})$ resulting from $\frac{\partial\mathbf{v}}{\partial x} = 0$.  Expanding Eq. (\ref{eq:F_x}) in terms of partial derivatives and rearranging gives 
\begin{equation}
F_x = q\left(-\frac{\partial A_x}{\partial t} - \frac{\partial \Phi}{\partial x}\right) + q v_y \left(\frac{\partial A_y}{\partial x}-\frac{\partial A_x}{\partial y}\right) -q v_z\left(\frac{\partial A_x}{\partial z} - \frac{\partial A_z}{\partial x}\right).
\label{eq:Lorentz_x}
\end{equation}
\noindent This is clearly the x-component of the Lorentz force law with the quantities in parenthesis being $E_x$, $B_z$, and $B_y$, respectively. 

Equation (\ref{eq:Lorentz_x}) can be expressed somewhat simpler in relativistic 4-vector or tensor notation.  Four-vectors have one \textit{temporal} and three \textit{spatial} components and are extremely useful for how they transform under a Lorentz transformation.  For example the position 4-vector $(ct,x,y,z)$  = $(ct,\mathbf{x})$ will transform to $(ct',\mathbf{x'})$, yet the scalar product $(ct)^2 - \mathbf{x}\cdot\mathbf{x}$ = $(ct')^2 - \mathbf{x'}\cdot\mathbf{x'}$  remains the same for any Lorentz transformation to any inertial coordinate system.  This invariant and its relationship to the Lorentz transformation is similar to the dot product and its relationship to rotation.  

The negative sign in the scalar product is dealt with by introducing two types of vectors that are related by flipping the sign of the spatial components.  The components of a contravariant 4-vector are represented by superscripts, e.g. $x^\mu$ where $(x^0,x^1,x^2,x^3)$ =  $(ct,x,y,z)$.  The components of a covariant 4-vector are represented by subscripts, e.g. $x_{\mu}$ represents $(ct,-x,-y,-z)$.  The scalar product then is the product of one covariant and one contravariant vector (the order is immaterial) and is represented as $x^\mu y_\mu$.  (Here, and for the rest of the paper, we use the summation notation where two repeated indexes in a product---one covariant and one contravariant---implies a sum over the indices.)  The usefulness of the scalar product is that if $x^\mu$ and $y^\mu$ are 4-vectors (in other words $x^\mu x_\mu$ and $y^\mu y_\mu$ are invariant) then the scalar product $x^\mu y_\mu$ is also invariant.     

Relativistic equations are expressed in their simplest form in terms of 4-vectors (and their scalar and tensor counterparts).  Important examples of contravariant 4-vectors include proper velocity $\eta^\mu = \gamma(c,\mathbf{v})$ where $\gamma$ is the relativity factor $\gamma = \left(1-(v/c)^2 \right)^{-1/2}$, 4-momentum $p^\mu = m\eta ^\mu$ where m is the rest mass, the 4-potential $A^\mu$ = $(\Phi/c,\mathbf{A})$, and the 4-current density $J^\mu = (c\rho,\mathbf{J})$, where $\rho$ and $\mathbf{J}$ are the charge and current densities.  Covariant and contravariant 4-vector derivatives have the negative sign reversed for the spatial part. The covariant derivative $\partial_\mu$ is $(\frac{1}{c}\frac{\partial}{\partial t},\nabla)$ and the contravariant derivative $\partial^\mu$ is $(\frac{1}{c}\frac{\partial}{\partial t},-\nabla)$ where $\nabla$ is the gradient in Cartesian coordinates.  

The Lorentz force equation (\ref{eq:Lorentz_x}) becomes in the tensor notation (after multiplying by $\gamma$):
\begin{equation}
K^\mu = q\eta_\nu \left(\partial^\mu A^\nu - \partial^\nu A^\mu \right),
\label{eq:Lorentz^mu}
\end{equation}
\noindent where $K^\mu$ is the Minkowski force whose spatial part $\mathbf{K}$ is $\gamma\mathbf{F}$.  (The temporal portion of the Minkowski force is $\gamma \frac{dW}{dt}$, where $\frac{dW}{dt}$ is the applied power.)  This can be simplified further by defining the anti-symmetric field 4-tensor $F^{\mu\nu} \equiv \left(\partial^\mu A^\nu - \partial^\nu A^\mu \right)$.  Evaluating $F^{\mu\nu}$ we see that its 6 unique values are the x, y, and z components of $\pm \frac{\mathbf{E}}{c}$ and $\pm \mathbf{B}$. 

\section{Gauge Invariance}

Equation (\ref{eq:Lorentz^mu}) introduces a difficulty with $A^\mu$. If we let $A^\mu\rightarrow \tilde{A}^\mu = A^\mu + \partial^\mu f$ where $f$ is \emph{any} arbitrary differentiable function it will lead to the exact same field tensor $F^{\mu\nu}$ as the original $A^\mu$.  This property is known as gauge invariance.  And, at first glance, it seems to be a major problem in interpreting $A^\mu$ as a physical quantity since $A^\mu$ is arbitrary to whole classes of functions!

It is important to note that this affects not just electrodynamics but all representations of energy and momentum because the derivation of Eq. (\ref{eq:Lorentz^mu}) is more general than just electrodynamics. If we absorb q into $A^\mu$ then Eq. (\ref{eq:Lorentz^mu}) applies to all systems that have minimal coupling.  This includes conservative fields as a special case where the spatial components of $A^\mu$ are zero.  

It is well understood---though not well known outside gauge theory---that gauge invariance is a general property of classical mechanics.\cite{Konopinski:CM}  Consider the well-known transformation of a Lagrangian $L\rightarrow L' = L - \frac{df}{dt}$, where f is any function of x and t.  This transformed Lagrangian $L'$ produces the same equations of motion as the original $L$.  (Recall that the Euler-Lagrange equations come from finding the path that minimizes the action $S=\int L dt$.  Therefore, by the fundamental law of calculus, the difference between the $S$ and $S'$ is a constant and cannot affect the path of least action.)  

But, transforming $L$ changes the momentum $\mathbf{p}$ and energy $H$ of the system.  Using the appropriate chain rule for the full derivative, $\frac{df}{dt} = \frac{\partial f}{\partial t} + \sum \frac{\partial f}{\partial x_{i}}\frac{dx_i}{dt}$, where the sum is over the three spatial components (x,y,z), we calculate $p'_{i}$ and $H'$ in the usual way:

\begin{eqnarray}
p'_{i} = \frac{\partial L'}{\partial \dot{x}_{i}} = p_{i} - \frac{\partial f}{\partial x_{i}} \nonumber \\
H' = \sum p'_{i}\dot{x}_i - L' = H + \sum\left(-\frac{\partial f}{\partial x_i}\right)\dot{x}_{i} -\left(-\frac{df}{dt}\right) = H + \frac{\partial f}{\partial t}, 
\label{Lagrange calculation of p and H}
\end{eqnarray}  

\noindent which are the equations for gauge invariance.  In other words the new $(H',\mathbf{p'})$ have the same equations of motion as the original $(H,\mathbf{p})$ for gauge transformations. 
  
The physical meaning of gauge invariance can be made clearer with an example reminiscent of the equivalence principle.  Imagine a rocket which when observed from an inertial reference frame has a constant acceleration $\mathbf{a}$.\cite{Goldstein}  An observer inside the rocket will clearly feel as if he is continually pulled downward. Observing a ball falling from someplace near to the nose of the rocket, he would conclude that there is a potential energy $V = may$ where m is the mass of the ball and y is its height.  The observer in the inertial reference frame, though, would insist that there was no downward force at all and definitely no potential energy.  She would see the floor accelerating into the ball, not the other way around. She would claim that the observer inside the rocket sees in \textit{his} reference frame a ball that is gaining momentum $=$ $-mat$. Defining a potential momentum $A$ in the same manner as potential energy is defined she could just as easily say that $A = +mat$ to keep momentum conserved.  

We could do the same in a different accelerating reference system in which case there would be another set of $(V,\mathbf{A})$ that are valid for a person on a rocket and produce the correct equations of motion in his reference system. In principle, the same should work for \textit{any} reference system, even one that varies with both position and time.  An observer in this frame will note that the person on the rocket sees a different set of $(V,\mathbf{A})$ that are now space and time dependent.  

Again, it is important to note, that this it \textit{not} a change of reference system.  All of these sets of $(V,\mathbf{A})$ are valid for the \textit{rocket's} reference frame.  All of these observers agree that the observer on the rocket sees a ball that has a Hamiltonian of the form of Eq. (\ref{eq:H(P,x)}) and has a force law of the form of Eq. (\ref{eq:Lorentz^mu}).  What they don't agree on is how the rocket observer should interpret $A^{\mu}$.  Gauge invariance reflects the fact that the motion of an object (for a set reference frame) should not depend on how you observe it.  Gauge invariance is a \textit{necessary} part of all of physics and is one of the cornerstones of all advanced theories of motion including classical mechanics, quantum mechanics, gauge theory, and general relativity.       

\section{Maxwell's Equations}

With the derivation of the Lorentz force complete, we turn our attention to Maxwell's equations.  There are many excellent derivations of Maxwell's equations starting from a variety of different assumptions.  See references \cite{AmJPhys,Jefimenko,Heras,Frisch&Wilets,Jackson2} for a small sampling.  Frisch and Wilets\cite{Frisch&Wilets} do a particularly good job of not only deriving the equations but listing the important ingredients that are necessary for its derivation.  

Here we will attempt to do the same but from the perspective of the potentials.  We will show that there are five relatively independent and necessary conditions underlying electrodynamics: 

\begin{enumerate}
\item \textit{Minimal Coupling}: $\mathbf{p}\rightarrow\mathbf{p}-q\mathbf{A}$,
\item \textit{Gauge Invariance}: $\partial_{\mu}A^{\mu} = 0$,
\item \textit{The 4-potential is carried by massless particles}: $\partial_{\nu}\partial^{\nu}A^{\mu} = 0$ in the absence of charge,
\item \textit{The 4-potential is directly proportional to 4-current density that created it}: $A^{\mu} \propto J^{\mu}$ 
\item \textit{Conservation of Charge}: $\partial_{\mu}J^{\mu} = 0$.
\end{enumerate} 

We have already used the first postulate to derive the Lorentz force law and have shown the necessity of gauge invariance (the second postulate).  The third postulate is due to massless photons mediating the electromagnetic force. The fourth postulate is necessary to derive the inhomogeneous Maxwell's equations and is supported by its simplicity and that it produces the correct field equations.  The final postulate is local charge conservation.

These postulates conveniently separate the electromagnetic interaction into three separate processes.  The fourth postulate represents how charged particles generates a 4-potential. The third postulate dictates how the 4-potential traverses from the source. (In this case it is carried by massless ``non-interacting'' photons.)  The first postulate represents how a charged particle reacts to the 4-potential it receives.  Finally, the second and the fifth postulates represent important additional restrictions.  

These are the assumptions that are needed to derive all of electrodynamics.  Furthermore, it is straight-forward to vary these postulates to model other forces.  For example, giving the photon a mass will alter the third postulate and lead to the Yukawa potential.

\subsection{The Homogeneous Maxwell's Equations}

The four homogeneous Maxwell equations are due to the minimal coupling condition although they are hidden in the definition of $F^{\mu\nu} = \left(\partial ^\mu A^\nu - \partial ^\nu A^\mu \right)$.  When written out in terms of spatial and time derivatives $F^{\mu\nu}$ includes gradients as well as curls and time derivatives of the components of $A^\mu$ in a simple form.  Therefore, we might expect there to be relationships in the derivatives of $F^{\mu\nu}$ similar to 
\begin{eqnarray}
\nabla \cdot \left(\nabla\times\mathbf{A}\right)= 0 = \frac{\partial}{\partial x_i} \left[ \epsilon_{ijk} \left( \frac{\partial}{\partial x_j}A_k \right) \right] \nonumber \\
\nabla\times \left(\nabla \Phi \right) = 0 = \frac {\partial}{\partial x_i}\left[\epsilon_{ijk} \left(\frac{\partial}{\partial x_j} \Phi \right) \right].
\label{eq:hmm}
\end{eqnarray}
Here, the Levi-Civita symbol is defined by $\epsilon_{ijk} = 0$ if any two of i,j,k are the same, $\epsilon_{ijk} = 1$ for all even permutations of 123, and $\epsilon_{ijk} = -1$ for all odd permutations of 123. It is straight-forward to show that there is a relationshipe for $F^{\mu\nu}$ similar to Eq. (\ref{eq:hmm}),
\begin{equation}
\partial _\mu \frac{1}{2}\epsilon ^{\mu\nu\rho\sigma}F_{\rho\sigma} = 0,
\label{eq:MaxwellHomogeneous}
\end{equation} 
\noindent that is valid as long as $A^\mu$ is differentiable.  (This is essentially due to the fact that partial derivatives commute for differentiable functions.) Here, the doubly covariant $F_{\rho\sigma}$ is obtained by changing the signs of the components of $F^{\mu\nu}$ that have both a spatial and temporal part, for example $F^{01}$ or $F^{30}$ but not $F^{31}$.  We also extend the Levi-Civita symbol to four dimensions in the expected way.  Evaluating Eq. (\ref{eq:MaxwellHomogeneous}) leads to the four homogeneous Maxwell's equations. This demonstrates that the homogeneous equations are valid for any system that acts on a particle with a Hamiltonian of the form of Eq. (\ref{eq:H(P,x)}). 

\subsection{The Inhomogeneous Maxwell's Equations}
In order to simplify the more complicated derivation of the inhomogeneous Maxwell's equations we split the derivation into two parts.  First we derive the form of the equation for regions where there are neither charges nor currents, $J^\mu = 0$ using the third postulate. Then we add a term proportional to $J^{\mu}$ in accordance with postulate 4 to account for the sources creating the 4-potentials.  

We start by examining the particles (photons) that are assumed to be mediating the electromagnetic force.  The relation between the relativistic energy and momentum of a particle having mass $m$ is given by  Eq. (\ref{eq:p^2}).  For a free particle (such as a photon at a location at which there is no 4-current) the total momentum and energy $(H,\mathbf{P})$ equals the relativistic energy and momentum $(E,\mathbf{p})$, respectively.  Substituting the appropriate quantum mechanical operators for $\mathbf{p}$ and $H$ we see that a particle with mass $m$ must have a wave function $\psi$ that satisfies the differential equation $\left(\frac{1}{c^2}\frac{\partial^2}{\partial t^2} - \nabla^2 + \left(\frac{mc}{\hbar}\right)^2\right)\psi = 0$, or in covariant notation $\left(\partial_\lambda\partial^\lambda + \left(\frac{mc}{\hbar}\right)^2\right)\psi = 0$.  Since the 4-potential is mediated by massless non-interacting photons (postulate 3), it is reasonable to assume that the 4-potential follows this same equation (with the photon having zero rest mass):

\begin{equation}
\partial_\lambda\partial^\lambda A^\mu = 0,
\label{waveA}
\end{equation}

\noindent valid for regions where $J^\mu = 0$. Note that extending this model to give the photon a small mass is straight-forward.  This equation represents a wave equation for the $A^\mu$. It should also be noted that in choosing our operators for $H$ and $p$ we have chosen a particular gauge.  We will have to enforce gauge invariance later.  
  
We incorporate the source term $J^\mu = (c\rho, \mathbf{J})$ by noting the 4-vector nature of $J^\mu$ and that the fields are linear in $J^\mu$ by postulate 4.  The simplest 4-vector equation (linear in $J^\mu$) that reduces to Eq. (\ref{waveA}) for $J^\mu = 0$ is 
\begin{equation}
\partial _\lambda \partial^\lambda \tilde{A}^\mu = \mu_o J^\mu,
\label{eq:MaxwellLorentzGauge}
\end{equation} 
\noindent where $\mu _0$ is a constant and we have marked $\tilde{A}^\mu$ with a tilde to remind us that this equation is for a particular gauge.  

To proceed we need to determine the particular gauge of $\tilde{A}^\mu$ such that Eq. (\ref{eq:MaxwellLorentzGauge}) leads to the conservation of charge (postulate 5),
\begin{equation}
\partial _\mu J^\mu = 0.
\label{eq:continuity}
\end{equation}
\noindent The solution to Eq. (\ref{eq:MaxwellLorentzGauge}) for the boundary condition that $A^\mu = 0$ at infinity is well known\cite{Jackson3},
\begin{equation}
\tilde{A} ^\mu = \frac{\mu _0}{c} \int \frac{J^\mu(ct',x',y',z')\delta\left(R-c(t-t')\right)}{R} d(ct')dx'dy'dz',
\label{eq:A(J)}
\end{equation}
\noindent where $R = \sqrt{(x-x')^2+(y-y')^2 + (z-z')^2}$ and $\delta$ is the Dirac delta function and the integral is over all space.  To determine the gauge of $\tilde{A}^\mu$ we need to determine the value of 
\begin{equation}
\partial _\mu \tilde{A} ^\mu  = \frac{\mu _0}{c} \int J^\mu(ct',x',y',z') \partial _\mu \frac{ \delta\left(R-c(t-t')\right)}{R} d(ct')dx'dy'dz'.
\label{eq:A-gauge}
\end{equation}
\noindent Using the symmetry between the primed and unprimed coordinates $\partial _\mu'\left[ \frac{\delta \left(R - c(t-t')\right)}{R}\right]$ $=$ $-\partial _\mu \left[\frac{\delta \left(R - c(t-t')\right)}{R}\right]$, we switch the derivative to the prime coordinates and then integrate by parts.  Using the product rule for differentiation we find that $\partial _\mu ' \left[J^\mu \frac{ \delta\left(R-c(t-t')\right)}{R}\right]$ $=$ $J^\mu\partial _\mu ' \left[\frac{ \delta\left(R-c(t-t')\right)}{R}\right]$ $+$ $\partial _\mu ' \left[J^\mu\right] \frac{ \delta\left(R-c(t-t')\right)}{R}$ = $J^\mu\partial _\mu ' \left[\frac{ \delta\left(R-c(t-t')\right)}{R}\right]$ where the second term is zero in the middle equation because of postulate 5, Eq. (\ref{eq:continuity}).  Therefore,
\begin{equation}
\partial _\mu \tilde{A}^\mu = \frac{\mu _0}{c} \int \partial _\mu ' \left[\frac{J^\mu(ct';\mathbf{x})\delta \left(R - c(t-t')\right)}{R} \right]d^4x'. 
\label{eq:A-gauge2}
\end{equation}
\noindent This volume integral evaluates as a surface integral in 4-space of the argument of the differential by an extension of the divergence theorem, where the surface is at plus or minus infinity in space and time.  As long as $J^\mu$ is localized such that it goes to zero faster than $1/R$ in the limit that $R$ goes to infinity then the value of Eq. ({\ref{eq:A-gauge2}) $= 0$.  Therefore the gauge of $\tilde{A}^\mu$ in Eq. (\ref{eq:MaxwellLorentzGauge}) is the Lorentz gauge $\partial _\mu \tilde{A} ^\mu = 0$.  

Generalizing Eq. (\ref{eq:MaxwellLorentzGauge}) to an arbitrary gauge is straight-forward since $A^\mu = \tilde{A} ^\mu + \partial ^\mu f$ where $A^\mu$ is the potential in an arbitrary gauge determined by $f$.  Plugging $\tilde{A} ^\mu = A^\mu - \partial ^\mu f$ into Eq. (\ref{eq:MaxwellLorentzGauge}) (using $\nu$ instead of $\lambda$) and using $\partial_\nu A^\nu = \partial _\nu \partial ^\nu f$ gives
\begin{equation}
\partial _\nu \left[\partial ^\nu A^\mu - \partial ^\mu A^\nu \right] = \mu _0 J^\mu = - \partial_\nu F^{\mu\nu}.
\label{eq:Maxwell}
\end{equation}
\noindent Using the definition of the field tensor, it is easily verified that this leads to the four inhomogeneous Maxwell equations.\cite{Griffiths2}

\section{Conclusions}
Eighty years after quantum mechanics has shown that the potentials are important fields of electrodynamics corresponding to the 4-momentum per unit charge transferred by the fields, the potentials still don't get the respect they deserve in the undergraduate electricity and magnetism course.  This is largely due to gauge invariance.  Here, we have shown that the Lorentz force law and the homogeneous Maxwell's equations of electrodynamics are a natural consequence of the 4-potential representing the potential energy and momentum per unit charge.  Furthermore, we have derived the entire set of Maxwell's equation from the 4-potential in a fully transparent way, explicitly showing all the necessary assumptions that are built into the equations.  As part of this discussion we have demonstrated that the phenomenon of gauge invariance is a necessary and important property of energy and momentum that affects all systems and therefore the scalar and vector potentials should be seen as being just as `physical' as energy and momentum.

%\appendix

%\section{The First Appendix}  The appendix fragment is used only once. Subsequent appendices can be created using the Section Section/Body Tag.
\end{document}